\documentclass[12pt]{iopart}

\begin{document}

\title{The quark vacuum}

\author{A. Miranda}
\address{Department of Physics and Astronomy, University of Aarhus,\\
DK-8000 Aarhus C, Denmark.\\
E-mail: miranda@phys.au.dk}

\begin{abstract}
We conjecture on the structure of the quark vacuum from a viewpoint somewhat
different from, but perhaps supplementary to, standard philosophies.

Using a rather simple dynamical Hamiltonian model, vacuum excitations
carrying helicities 0 and $\pm 1$ are discussed in connection with the
dynamical stability of solutions. We speculate on how the rest masses of the
light mesons $\pi ^{0}$, $\eta ^{0}$and $\eta ^{0^{\prime }}$ could be
related to these excitations.

\end{abstract}


\maketitle

\section{Introduction}

We assume in this paper that the physical vacuum, which has acquired a place
of honour in contemporary physics because of the great empirical success of
the Standard Model, is a physical medium that can be probed just like any
other physical medium \cite{1}. This idea has, of course, a long history behind it.
In modern physics, it first emerged as a possibility, suggested by detailed
QED studies. It seems, however, that Nature has provided us with many
alternative ways of empirically testing this idea . We are referring in fact
to the physics of chiral and flavour symmetries and symmetry breaking in the
world of quarks and leptons, as assumed and described by the Standard Model
\cite{2}. In contemporary physics, this is usually tackled by relying on the
generally very successful methodology of Effective Field Theories \cite{2,3,4,5}.

In this paper, we concentrate on one of the best hints we have about the
nature of this presumed medium, through the structure of the lightest
self-conjugate mesons. This subject has of course been discussed in detail
in the last 40 years or so, but we believe that the case is not definitely
closed as yet. The general question this paper studies is:
if the vacuum indeed is a mostly
unknown physical medium, then how do approximate \textit{ansatzes }about
the nature of this presumed medium manifest themselves on key properties of
the predicted particle spectrum.

We begin by reformulating this question in terms of Dirac's Quantum Field
Theory ``without dead wood'' (\cite{6}; Appendix I). This means that questions
about symmetry, spontaneous symmetry breaking and chiral anomalies are
formulated in a way that differ somewhat from the standard one \cite{2,3,4,5}. The
main technical points are, first we abandon the notion of wave-functionals
and vacuum expectation values; this implies among other things, in a
perturbative setting and at any stage of the development of the theory, the
automatic non-existence of such irrelevant objects as unconnected vac-to-vac
diagrams. This non-trivial streamlining of Quantum Field Theory should of
course lead to the same results as standard formulations, in so far as these
agree with experiments. However, the interpretation differs.

We further emphasise that we always work within momentum space. This feature
we share with pure S-matrix theory, although we have no S-matrix here.

Specialized principles that we assume are mutually compatible and that we
attempt to obey are:

a) essential ideas of the empirically rather successful quark model for
mesons \cite{3,4}, which are defined only in the so-called confinement phase;

b) the reliability of the QCD assumptions (especially chiral-flavour
symmetries) for the physics at very short distances (i.e. d$\ll \Lambda
_{QCD}^{-1}$ with $\Lambda _{QCD}$ of the order of 0.2 - 0.3 GeV)\cite{2,3,4};

c) the presumed moderate reliability of ENJL-like models for the relevant
physics of the medium distance quark dynamics (i.e. d $>\Lambda _{\chi
}^{-1} $ but d$<\Lambda _{QCD}^{-1}$ with $\Lambda _{\chi }$ of the order of
1GeV) \cite{7}. We do not actually use this particular model.

A short preview of this paper would be in order.

In Sections 2 and 3 we present the mathematical machinery used in this and
subsequent papers. This is based on what Dirac recommended in his later
years \cite{6}, and is eminently suitable to our purposes. It may be rather
unfamiliar to many particle theoreticians, but the material in App.A may
help.

In Section 4 the master equations previously established are applied to the
so-called \textit{current quarks,} which are our fundamental (Weyl) \textit{
q-numbers} (Dirac's terminology,\cite{6}). Starting with massless particles in
our Model Hamiltonian, we discuss three cases, viz. all quarks massless
(case A); next , massless (u,d) quarks, massive s-quark (case B); and
finally, three massive species, but obeying the hierarchy $0<m_{u}<m_{d}\ll
m_{s}$(case C).

We show that all quarks become massive through pure quantum effects. We
study the results of self-energy corrections to the quarks energy-momentum
relationship, and verify the robustness (stability) of the vacuum exhibiting
such effects. Particular features of this calculation (which to some extent
are common to all cases) are:

(i) automatic output of \textit{massive} Dirac particles, even with input
Weyl fermions;

(ii) the emergence of Bose-like states (again in Dirac's sense of this word
,Ref 6), one of these a massless state (Goldstone,case A), or a nearly
massless one (Pseudo-Goldstone; cases B and C);

(iii) the emergence of another massive highly coherent superposition of $u%
\bar{u},d\bar{d}$ and $s\bar{s}.$

The relationship of these vacuum excitations to $\pi ^{0},\eta ^{0}$
and $\eta ^{0\prime}$ is conjectured upon in App.B.

We do not know at the moment how to formulate the problem of confinement in
our approach (at least as discussed in the literature), let alone find a
solution for it. We have therefore completely by-passed this issue. All
gauge fields of the Standard Model (that are assumed to couple to our
quarks) provide only some sort of an implicit unanalysed background.

Questions having to do with the electroweak couplings of quarks, although
very important to our views, shall be discused in a future paper.

\section{The theoretical framework}

The theory used in this paper claims to be yet another example of
how to construct a ``field theory without dead wood'' in practice \cite{6}. Some
advantages of this more logical (and simpler!) approach to field
theoretical physical problems (in practice treated by standard perturbative
methods) were spelled out by Dirac in his later years \cite{6}.

\subsection{The Model Hamiltonian}

We need an effective field theoretic Hamiltonian \cite{2,3,4}, or a Model
Hamiltonian at the level of QCD-quarks that must be restricted by the
following conditions:

1) Baryon number and charge conservation ;

2) CPT-invariance (we further assume CP-invariance);

3) share with QCD its global colour, flavour and chiral symmetries (we
bypass the confinement issue);

4)have a minimal number of special assumptions and parametrizations.

The S-matrix at the quark level is not defined, but this effective
Hamiltonian should ensure that one eventually ends up with an \textit{%
hadronic\ }S-matrix with the usual properties \cite{2}.

There are of course in principle an infinity of ``microscopic'' quark
Hamiltonians satisfying these conditions appropriate to the vacuum phenomena
discussed in this paper, even if we restrict ourselves only to the first two
families and to at most quadrilinear terms as effective interactions.

Our model Hamiltonian for QCD-quarks is assumed to have the form:

\begin{equation}
\hat{H}=U_{B}+\hat{H}_{q}+\hat{H}_{ew}  \label{2.1A}
\end{equation}

\begin{equation}
\hat{H}_{q}=\hat{H}_{0q\bar{q}}+\hat{V}_{q\bar{q}}^{(1)}+\hat{V}_{q\bar{q}%
}^{(2)}  \label{2.1B}
\end{equation}

If $\Omega $ represents the quantization box volume, then $U_{B}/\Omega $ is
a ``background energy density '' as $\Omega \rightarrow \infty $, supposed
to absorb the infinite zero-point energy of the virtual Dirac-Weyl quark
seas.

$\hat{H}_{0q\bar{q}},\hat{V}_{q\bar{q}}^{(1)}$ $,\hat{V}_{q\bar{q}}^{(2)}$
and $\hat{H}_{ew}$ are pieces of our model Hamiltonian trying to capture
relevant key features of QCD-quarks and their assumed effective interactions
as they emerge within our model space $\mathcal{D}$.

In general we should include in $\hat{H}_{q}$ all relevant terms that are at
most quadrilinear in the preassigned fundamental fermion representations
\cite{7}. We conjecture, however , that the minimal relevant terms for trying to
understand the intricate chiral and flavour dynamics of the lightest quarks
can be restricted as follows:

\begin{equation}
\hat{H}_{0q\bar{q}}[\mathcal{D}]=\sum_{c,\mu }\epsilon _{\mu }[:\alpha
_{c\mu }^{+}\alpha _{c\mu }-\beta _{\bar{c}\bar{\mu}}\beta _{\bar{c}\bar{\mu}%
}^{+}:]  \label{2.2}
\end{equation}

and

\begin{equation}
\hat{V}_{q\bar{q}}^{(1)}[\mathcal{D}]=\sum_{c,\mu }x_{\mu }[:\alpha _{c\mu
}^{+}\beta _{\bar{c}\bar{\mu}}^{+}+\beta _{\bar{c}\bar{\mu}}\alpha _{c\mu }:]
\label{2.3}
\end{equation}

\begin{equation}
\hat{V}_{q\bar{q}}^{(2)}[\mathcal{D}]=\sum_{c,c^{\prime }}\sum_{\mu ,\nu \in 
\mathcal{D}}\sum_{\mu ^{\prime },\nu ^{\prime }\in \mathcal{D}}:\alpha
_{c\mu }^{+}\beta _{\bar{c}\bar{\mu}^{\prime }}^{+}<\mu \bar{\mu}^{\prime }|%
\hat{V}_{q\bar{q}}^{(2)}[\mathcal{D}]|\nu \bar{\nu}^{\prime }>\beta _{\bar{c}%
^{\prime }\bar{\nu}^{\prime }}\alpha _{c^{\prime }\nu }:  \label{2.4}
\end{equation}

We assume that the input q-numbers $\alpha _{c\mu }^{+}$ etc. are massless
colour triplet states carrying quantum numbers $c$ and $\mu \equiv (n,\vec{p}%
,\lambda )$ standing respectively for colour and (flavour, 3-momentum,
helicity) ; $\beta _{\bar{c}\bar{\mu}}^{+}$ are their antiparticle states,
here defined to mean their CP-conjugate states. The symbol :: stands for
``normal products'' to be presently specified .

The coefficients $x_{\mu }$ are part of the input parameters defining our
model space.

We thus assume that we are dealing with the flavour representations (with
usual assignements for $u,d,s$ quarks)

\begin{equation}
n=\{(\frac{1}{2},i_{3})\otimes (Q_{\frac{1}{2}},S_{\frac{1}{2}})\}\oplus
\{(0,0)\otimes (Q_{0},S_{0})\}  \label{2.5A}
\end{equation}

\[
n=1\longleftrightarrow (u,d) 
\]

\begin{equation}
u=(\frac{1}{2},+\frac{1}{2})\otimes (\frac{2}{3},0)\qquad d=(\frac{1}{2},-%
\frac{1}{2})\otimes (-\frac{1}{3},0)  \label{2.5B}
\end{equation}

\[
n=2\longleftrightarrow (s) 
\]

\begin{equation}
s=(0,0)\otimes (-\frac{1}{3},-1)  \label{2.5C}
\end{equation}

The various relevant model matrix elements (\ref{2.4}) are specified in Section 5.
The summation symbols stand for ordinary summations over discrete quantum
numbers and integrals over 3-momenta.

It is important to note that interactions among our input quasiparticles are
defined only within a momentum space $\mathcal{D}$ extending up to about $%
\Lambda _{\chi }$ from zero momentum (Section 5).

Colour degrees of freedom are not dynamically involved in this model. We
have thus nothing to say about confinement as generally understood in the
literature.

\section{The approximation scheme}

\subsection{Fermions}

The general idea is to introduce a succesion of canonical transformations
designed to diagonalize as much of this Hamiltonian as possible. The
hopefully small and neglected residual contributions can be added
perturbatively, if necessary.

Thus, let us begin with the fundamental q-numbers themselves:

\begin{equation}
N_{c\mu }^{^{\prime }}\equiv \left( 
\begin{array}{l}
b_{c\mu } \\ 
d_{\bar{c}\bar{\mu}}^{+}
\end{array}
\right) =\left( 
\begin{array}{ll}
u_{\mu } & v_{\mu } \\ 
-v_{\mu }^{*} & u_{\mu }^{*}
\end{array}
\right) \left( 
\begin{array}{l}
\alpha _{c\mu } \\ 
\beta _{\bar{c}\bar{\mu}}^{+}
\end{array}
\right) =  \label{3.1}
\end{equation}

\[
=\mathbf{U}_{\mu }N_{c\mu } 
\]

The folowing condition upon the transformation matrix is imposed, ensuring
its invertibility:

\begin{equation}
|u_{\mu }|^{2}+|v_{\mu }|^{2}=1  \label{3.2}
\end{equation}

We furthermore assume (without loss of generality) that these (variational)
parameters $u_{\mu }$ and $v_{\mu }$ are constrained to satisfy the
conditions

\begin{equation}
u_{\mu }=u_{\bar{\mu}}=u_{\mu }^{*}\equiv \cos \frac{\theta _{\mu }}{2}%
\qquad v_{\mu }=v_{\bar{\mu}}=v_{\mu }^{*}\equiv \sin \frac{\theta _{\mu }}{2%
}  \label{3.3}
\end{equation}

hence

\begin{equation}
\mathbf{U}_{\mu }=\exp (i\frac{\theta _{\mu }}{2}\mathbf{\tau }_{2}) 
\label{3.4}
\end{equation}

where $\mathbf{\tau }_{2}$ is one of Pauli matrices.

Our approximation scheme is basically as follows. The q-numbers $N_{c\mu
}^{^{\prime }}$ generated by the ``Dirac-Bogoliubov rotation'' (\ref{3.4}) should
be \textit{stationary states} (to borrow Dirac's terminology \cite{6} , see also
Appendix A)) :

\begin{equation}
\lbrack \hat{H},b_{\mu }^{+}]=E_{\mu }b_{\mu }^{+}  \label{3.5A}
\end{equation}

\begin{equation}
\lbrack \hat{H},d_{\bar{\mu}}^{+}]=E_{\mu }d_{\bar{\mu}}^{+}  \label{3.5B}
\end{equation}

We expect that

\begin{equation}
\hat{H}_{q}=U_{0}+\hat{H}_{0}^{(I)}+\hat{V}_{res}^{(I)}(b^{+},b,d^{+},d) 
\label{3.6A}
\end{equation}

\begin{equation}
\hat{H}_{0}^{(I)}=\sum_{c}\sum_{\mu }E_{\mu }[b_{c\mu }^{+}b_{c\mu }-d_{\bar{%
c}\bar{\mu}}d_{\bar{c}\bar{\mu}}^{+}]  \label{3.6B}
\end{equation}

with $\hat{V}_{res}^{(I)}$ being our first ``residual'', which is at least
quadrilinear. We choose the `` Dirac-Bogoliubov rotation angles'' $\theta
_{\mu }$ so that this residual interaction is as small as possible. We
should, however, next check whether or not these residuals could destabilize
our zeroth order solutions. We may then continue with this process, pulling
more and more out of this $\hat{V}_{res}^{(I)}$ until ordinary perturbation
theory eventually may take over, if desired.

In order to reach this goal, we first compute the commutators , \textit{%
linearize the resulting equations }and look for non-trivial solutions. Using
the previously defined effective interaction and putting

\[
G_{n\lambda }(\vec{p})= 
\]

\begin{equation}
-\sum_{c^{\prime }}\sum_{n^{\prime }}\sum_{\lambda ^{\prime }}\int d^{3}\vec{%
p}^{\prime }<n^{\prime }\bar{n}^{\prime };\lambda ^{\prime }\vec{p}^{\prime
}|\hat{V}|n\bar{n};\lambda \vec{p}>u_{n^{\prime }p^{\prime }}v_{n^{\prime
}p^{\prime }}  \label{3.7}
\end{equation}

\[
F_{n\lambda }(\vec{p})\equiv 
\]

\begin{equation}
\equiv \sum_{c^{\prime }}\sum_{n^{\prime }}\sum_{\lambda ^{\prime }}\int
d^{3}\vec{p}^{\prime }<n^{\prime }\bar{n}^{\prime };\lambda ^{\prime }\vec{p}%
^{\prime }|\hat{V}|n\bar{n};\lambda \vec{p}>f_{n^{\prime }\lambda ^{\prime
}n\lambda }(\vec{p},\vec{p})  \label{3.8A}
\end{equation}

where

\begin{equation}
f_{n^{\prime }\lambda ^{\prime }n\lambda }(\vec{p},\vec{p})\equiv <\alpha
_{n^{\prime }\vec{p}^{\prime }\lambda ^{\prime }}^{+}\alpha _{n\vec{p}%
\lambda }+\beta _{\bar{n}^{\prime }\vec{p}^{\prime }\lambda ^{\prime
}}^{+}\beta _{\bar{n}\vec{p}\lambda }>  \label{3.8B}
\end{equation}

  we find Eq.(\ref{3.7}) which will be referred to as ``the gap
equation''. Eqs.(\ref{3.8A},\ref{3.8B}) represent corrections to the single quark/antiquark
energies.

We find that the condition for non-trivial solutions is:

\begin{equation}
\left| 
\begin{array}{ll}
|\vec{p}|+F_{n\lambda }(\vec{p})-E_{n\lambda }(p) & \qquad \quad \epsilon
_{n}+G_{n\lambda }(\vec{p}) \\ 
\qquad \quad \epsilon _{n}+G_{n\lambda }(\vec{p}) & -\{|\vec{p}%
|+F_{n\lambda }(\vec{p})\}-E_{n\lambda }(p)
\end{array}
\right| =0  \label{3.9}
\end{equation}

and so

\begin{equation}
E_{n\lambda }^{2}(p)=|\vec{p}|^{2}+\Delta _{n\lambda }^{2}(\vec{p}) 
\label{3.10}
\end{equation}

where

\begin{equation}
\Delta _{n\lambda }^{2}(\vec{p})\equiv 2|\vec{p}|F_{n\lambda }(\vec{p}%
)+F_{n\lambda }^{2}(\vec{p})+[\epsilon _{n}+G_{n\lambda }(\vec{p})]^{2} 
\label{3.11}
\end{equation}

Furthermore

\begin{equation}
v_{n\lambda }^{2}(\vec{p})=\frac{1}{2}(1-\frac{\epsilon _{n\lambda }(\vec{%
p})}{E_{n\lambda }(\vec{p})})  \label{3.12}
\end{equation}

\begin{equation}
u_{n\lambda }^{2}(\vec{p})=\frac{1}{2}(1+\frac{\epsilon _{n\lambda }(\vec{%
p})}{E_{n\lambda }(\vec{p})})  \label{3.13}
\end{equation}

These are the first set of our master equations. Self-consistency requires
that non-trivial solutions of eqs(\ref{3.9}) must be positive definite.

Solutions are discussed in Section 5.

\subsection{Bosons}

We shall next consider another branch of flavourless boson-like vacuum
excitation modes. This should give some feeling about the dynamical
stability of the above vaccum solutions, at least in these particular
channels. We continue to work within the framework of linearization
procedures.

The Hamiltonian (\ref{3.6A}) and (\ref{3.6B}) is thus rewritten in the form:

\begin{equation}
\hat{H}[\mathcal{D}]=U_{0}+\sum_{c}\sum_{n}\sum_{\lambda }\int d^{3}\vec{p}%
E_{n}(p)(b_{cn\vec{p}\lambda }^{\dagger }b_{cn\vec{p}\lambda }+d_{\bar{c}%
\bar{n}\vec{p}\lambda }^{+}d_{\bar{c}\bar{n}\vec{p}\lambda })+  \label{3.14}
\end{equation}

\[
+\sum_{c,c^{\prime }}\sum_{n,n^{\prime }}\sum_{\lambda _{1},\lambda
_{2}}\sum_{\lambda _{3}\lambda _{4}}\int d^{3}\vec{k}^{\prime }\int d^{3}%
\vec{k}<n^{\prime }\vec{p}^{\prime }\lambda _{3};\bar{n}^{\prime }-\vec{p}%
^{\prime }\lambda _{4}|\hat{V}|n\vec{p}\lambda _{1};\bar{n}-\vec{p}\lambda
_{2}> 
\]

\[
:\alpha _{c^{\prime }n^{\prime }\vec{p}^{\prime }\lambda _{3}}^{+}\beta _{%
\bar{c}^{\prime }\bar{n}^{\prime }-\vec{p}^{\prime }\lambda _{4}}^{+}\beta _{%
\bar{c}\bar{n}-\vec{p}\lambda _{2}}\alpha _{cn\vec{p}\lambda _{1}}: 
\]

  We look for \textit{stationary solutions} of the equation of
motion:

\begin{equation}
i\frac{\partial }{\partial t}\mathbf{\hat{O}}=[\hat{H}[\mathcal{D}],\mathbf{%
\hat{O}}]  \label{3.15}
\end{equation}

A key condition for obtaining such stationary solutions is of course that
they should obey the conservation laws associated with the symmetries of the
effective (model) Hamiltonian $\hat{H}[\mathcal{D}].$

We first look for solutions in terms of our primary composites:

\[
\mathbf{\hat{O}}=\mathbf{\hat{O}}(\Gamma ^{+},\Gamma ) 
\]

\begin{equation}
\Gamma _{\rho \bar{\rho}^{\prime }}^{+}(\vec{p})\equiv \frac{1}{\sqrt{3}}%
\sum_{c}b_{cf\vec{p}\lambda _{a}}^{+}d_{\bar{c}\bar{f}-\vec{p}\lambda
_{b}}^{+}  \label{3.16A}
\end{equation}

\begin{equation}
\Gamma _{\rho \bar{\rho}^{\prime }}(\vec{p})\equiv \frac{1}{\sqrt{3}}%
\sum_{c}d_{\bar{c}\bar{f}-\vec{p}\lambda _{b}}b_{cf\vec{p}\lambda _{a}} 
\label{3.16B}
\end{equation}

In order to ease the notation we conventioned that

\[
\rho \bar{\rho}^{\prime }\Longleftrightarrow f\lambda _{a}\lambda _{b}(\vec{p%
})\Longleftrightarrow r\qquad etc 
\]

Let us try the ansatz (appropriate momentum integrations understood)

\[
\mathbf{\hat{O}}\equiv \Gamma _{K}^{+}=\sum_{r}X_{Kr}\Gamma
_{r}^{+}-\sum_{r}Y_{Kr}\Gamma _{r}+ 
\]

\begin{equation}
+\sum_{r,m}a_{mr}\Gamma _{m}^{+}\Gamma _{r}^{+}+\sum_{r,m}b_{mr}\Gamma
_{m}^{+}\Gamma _{r}+\sum_{r,m}c_{mr}\Gamma _{r}\Gamma _{m}+...  \label{3.17}
\end{equation}

In this paper, approximations are limited to linearizing the equation of
motion. This is of course consistent with the basic assumption that our
zeroth order solutions are truly stable ones, and therefore non-linear
corrections are small and expected to contribute only to higher orders.
So,

\begin{equation}
\Gamma _{K}^{+}(\vec{p})=\sum_{r}\int_{\mathcal{D}}d^{3}\vec{p}^{\prime
}X_{Kr}(p,p^{\prime })\Gamma _{r}^{+}(\vec{p}^{\prime })-\sum_{r}\int_{%
\mathcal{D}}d^{3}\vec{p}^{\prime }Y_{Kr}(p,p^{\prime })\Gamma _{r}(\vec{p}%
^{\prime })  \label{3.18A}
\end{equation}

with the condition that

\begin{equation}
\sum_{r}\int_{\mathcal{D}}d^{3}\vec{p}^{\prime
}|X_{Kr}|^{2}-\sum_{r}\int_{D}d^{3}\vec{p}^{\prime }|Y_{Kr}|^{2}=1 
\label{3.18B}
\end{equation}

We start the procedure by trying to satisfy the equations

\begin{equation}
\lbrack \hat{H},\Gamma _{K}^{+}(\vec{p})]=\Omega _{K}(p)\Gamma _{K}^{+}(\vec{%
p})  \label{3.19}
\end{equation}

with the condition that

\begin{equation}
\Omega _{K}(p)\geq 0  \label{3.20}
\end{equation}

After some tedious but straightforward calculations we end up with our next
set of master equations to be used for studying stability questions and
conservation laws (in short-hand notation):

\begin{equation}
X_{K\rho \bar{\rho}^{\prime }}=\frac{1}{E_{\rho \bar{\rho}^{\prime
}}^{\prime }-\Omega _{K}}\sum_{\mu \bar{\mu}^{\prime }}\{-X_{K\mu \bar{\mu}%
^{\prime }}\mathcal{M}_{\mu \bar{\mu}^{\prime }\rho \bar{\rho}^{\prime
}}+Y_{K\mu \bar{\mu}^{\prime }}\mathcal{N}_{\mu \bar{\mu}^{\prime }\rho \bar{%
\rho}^{\prime }}\}\qquad \mathbf{(I)}  \label{3.21A}
\end{equation}

\begin{equation}
Y_{K\rho \bar{\rho}^{\prime }}=\frac{1}{E_{\rho \bar{\rho}^{\prime
}}^{\prime }+\Omega _{K}}\sum_{\mu \bar{\mu}^{\prime }}\{-Y_{K\mu \bar{\mu}%
^{\prime }}\mathcal{M}_{\mu \bar{\mu}^{\prime }\rho \bar{\rho}^{\prime
}}+X_{\mu \bar{\mu}^{\prime }}\mathcal{N}_{\mu \bar{\mu}^{\prime }\rho \bar{%
\rho}^{\prime }}\}\qquad \mathbf{(II)}  \label{3.21B}
\end{equation}

with the definitions:

\begin{equation}
\rho \bar{\rho}^{\prime }\Longleftrightarrow f\lambda _{a}\lambda _{b}(\vec{p%
})\qquad \mu \bar{\mu}^{\prime }\Longleftrightarrow n\lambda _{1}\lambda
_{2}(\vec{k})  \label{3.22A}
\end{equation}

\begin{equation}
E_{\rho \bar{\rho}^{\prime }\mu \bar{\mu}^{\prime }}^{\prime }\equiv 
\label{3.22B}
\end{equation}

\[
(E_{\mu }+E_{\bar{\mu}^{\prime }})\delta _{\mu \rho }\delta _{\bar{\mu}%
^{\prime }\bar{\rho}^{\prime }}+u_{\rho ^{\prime }}v_{\rho }<\mu \bar{\rho}|%
\hat{V}|\mu ^{\prime }\bar{\rho}^{\prime }>u_{\mu }v_{\mu ^{\prime
}}+u_{\rho }v_{\rho ^{\prime }}<\rho ^{\prime }\bar{\mu}^{\prime }|\hat{V}%
|\rho \bar{\mu}>u_{\mu ^{\prime }}v_{\mu } 
\]

\begin{equation}
\mathcal{M}_{\rho \bar{\rho}^{\prime }\mu \bar{\mu}^{\prime }}\equiv
3\{u_{\mu }u_{\mu ^{\prime }}<\mu \bar{\mu}^{\prime }|\hat{V}|\rho \bar{\rho}%
^{\prime }>u_{\rho }u_{\rho ^{\prime }}+  \label{3.22C}
\end{equation}

\[
+v_{\rho }v_{\rho ^{\prime }}<\rho ^{\prime }\bar{\rho}|\hat{V}|\mu ^{\prime
}\bar{\mu}>v_{\mu }v_{\mu ^{\prime }}\} 
\]

\begin{equation}
\mathcal{N}_{\rho \bar{\rho}^{\prime }\mu \bar{\mu}^{\prime }}\equiv
3\{v_{\mu }v_{\mu ^{\prime }}<\mu ^{\prime }\bar{\mu}|\hat{V}|\rho \bar{\rho}%
^{\prime }>u_{\rho }u_{\rho ^{\prime }}+  \label{3.22D}
\end{equation}

\[
+v_{\rho }v_{\rho ^{\prime }}<\rho ^{\prime }\bar{\rho}|\hat{V}|\mu \bar{\mu}%
^{\prime }>u_{\mu }u_{\mu ^{\prime }}\} 
\]

A comment on effects from couplings (\ref{3.14}) in the Hamiltonian (\textit{not}
included in Eqs.\ref{3.22A} and \ref{3.22B}) is in order. Within our linearized theory, it can
easily be verified that they do not contribute. Their effects, although
numerically unimportant within the context of our assumptions and
approximations, could be treated either by including the terms that were
left out and/or by selective summations, as used in standard perturbative
methods. In the latter case, the Feynman diagram rules must be constructed
in such a way that one can automatically perform precise corrections for any
double countings that may arise, due to this early emergence of free
point-like bosons in a pure fermionic theory.

It can also be easily verified that this scheme is a conserving
approximation, in the sense that it satisfies symmetries of the Hamiltonian,
and thus conservation laws. Thus, for example, let F be the generator of a
global gauge symmetry of the Hamiltonian. Should the approximate zero-th
order vacuum solution not be invariant under F, then a zero frequency
solution is automatically produced , in agreement with the Goldstone theorem
(further comments in Section 4). This approximation scheme respects
therefore the symmetries of the Hamiltonian to leading order of small
quantities (i.e. those that were left out in the calculations).

We seek non-trivial solutions for this system of equations.

\section{Numerical Results and Discussions}

We report in this paper the main numerical results based on the theory
presented in the previous sections. Explicit expressions for the matrix
elements (appropriate to our model space) are given in this section.

We assume:

\textbf{(i)}

\begin{equation}
<n^{\prime }\lambda ^{\prime };\bar{n}^{\prime }\lambda ^{\prime };\vec{p}%
^{\prime }|\hat{V}|n\lambda ;\bar{n}\lambda ;\vec{p}>\equiv <n^{\prime }\bar{%
n}^{\prime };\lambda ^{\prime };\vec{p}^{\prime }|\hat{V}|n\bar{n};\lambda ;%
\vec{p}>  \label{4.1}
\end{equation}

\begin{equation}
<n^{\prime }\bar{n}^{\prime };+;\vec{p}^{\prime }|\hat{V}|n\bar{n};+;\vec{p}%
>=<n^{\prime }\bar{n}^{\prime };+;\vec{p}^{\prime }|\hat{V}|n\bar{n};-;\vec{p%
}>=  \nonumber
\end{equation}

\begin{equation}
<n^{\prime }\bar{n}^{\prime };-;\vec{p}^{\prime }|\hat{V}|n\bar{n}+;\vec{p}%
>=<n^{\prime }\bar{n}^{\prime };-;\vec{p}^{\prime }|\hat{V}|n\bar{n};-;\vec{p%
}>=-\ \frac{|G|}{4\pi \Lambda _{\chi }^{2}}  \label{4.2}
\end{equation}

and

\textbf{(ii)}

\begin{equation}
<n^{\prime }+;\bar{n}^{\prime }-;\vec{p}^{\prime }|\hat{V}|n+;\bar{n}-;\vec{p%
}>=<n^{\prime }-;\bar{n}^{\prime }+;\vec{p}^{\prime }|\hat{V}|n-;\bar{n}+;%
\vec{p}>=\frac{g_{1}}{4\pi \Lambda _{\chi }^{2}}  \label{4.3}
\end{equation}

We have thus introduced a minimum of two further (dimensionless) parameters,
viz. $G$ and $g_{1}$. Note the explicit - sign on the rhs of definition \ref{4.2}.
As we shall see, this guarantees positive definite gaps .

This is supposed to represent effective colour singlet q\={q} interactions
at momenta below say $\Lambda _{\chi }\approx $ 1 GeV. This built-in limit
thereby acquires a physical significance in our model.

\subsection{Light quark rest masses}

The extensive analysis carried out by Leutwyler et al \cite{5} revealed that the
light quark masses obey the condition that their running masses in the $%
\stackrel{---}{MS}$ scheme at scale $\mu =1GeV$ must be

\begin{equation}
m_{u}=5.1\pm 0.9MeV\qquad m_{d}=9.3\pm 1.4MeV\qquad m_{s}=175\pm 25MeV 
\label{4.4}
\end{equation}

  Using eqs.(\ref{3.9}-\ref{3.13}) and the effective interaction (\ref{4.2}) we find
that:

\begin{equation}
\bar{E}_{n}^{2}(x)\equiv \bar{e}_{n}^{2}(x)+\bar{\Delta}_{n}^{2}  \label{4.5A}
\end{equation}

\begin{equation}
\bar{e}_{n}(x)=x-\frac{g}{6}\bar{v}_{n}^{2}(x)  \label{4.5B}
\end{equation}

\begin{equation}
\bar{\Delta}_{n}=x_{n}+gI(\bar{\Delta}_{n})  \label{4.5C}
\end{equation}

\begin{equation}
I(\bar{\Delta}_{n})=\sum_{n}\int_{0}^{1}x^{2}\bar{u}_{n}(x)\bar{v}_{n}(x)dx 
\label{4.5D}
\end{equation}

and

\begin{equation}
\bar{v}_{n}^{2}(x)=\frac{1}{2}(1-\frac{\bar{e}_{n}(x)}{\bar{E}_{n}(x)}) 
\label{4.6A}
\end{equation}

\begin{equation}
\bar{u}_{n}^{2}(x)=1-\bar{v}_{n}^{2}(x)  \label{4.6B}
\end{equation}

with the definitions

\begin{equation}
p=x\Lambda _{\chi }  \label{4.7}
\end{equation}

\[
\Delta _{n}=\bar{\Delta}_{n}\Lambda _{\chi } 
\]

\[
e_{n}(p)=\bar{e}_{n}(x)\Lambda _{\chi } 
\]

\[
E_{n}(p)=\bar{E}_{n}(x)\Lambda _{\chi } 
\]

\[
\epsilon _{n}=x_{n}\Lambda _{\chi } 
\]

\[
g=12|G|\Lambda _{\chi } 
\]

\[
\bar{\Delta}_{n}=\bar{\epsilon}_{n}+I(\bar{\Delta}_{n}) 
\]

2) The parameter space must be such that \cite{5}:

\begin{equation}
m_{s}(1GeV)=175\pm 25MeV  \label{4.8A}
\end{equation}

\begin{equation}
m_{d}(1GeV)=9.3\pm 1.4MeV  \label{4.8B}
\end{equation}

\begin{equation}
m_{u}(1GeV)=5.1\pm 0.9MeV  \label{4.8C}
\end{equation}

A sample of our numerical results can be seen in Fig.\ref{fig.1} through Fig.\ref{fig.8}. All
fits satisfy the additional constraints that $|f_{n}|<10^{-5}$ where

\begin{equation}
f_{n}=\bar{\epsilon}_{n}+I(\bar{\Delta}_{n})-\bar{\Delta}_{n}  \label{4.9}
\end{equation}

With these definitions we shall henceforth drop the bars on the symbols in
(\ref{4.7}).

\subsection{On the stability of solutions - normal modes}

We shall next examine the dynamical stability of the previous solution. A
reasonable criterion for stability,or near stability, is that all solutions $%
\Omega _{K\lambda }$ of the master equations (\ref{3.21A} and \ref{3.21B}) should be positive or
zero, assuming that the gap equation (\ref{4.5C}) is satisfied. The value zero
would imply that the chosen vacuum solutions are unstable according to this 
\textit{linearized} theory.

\subsubsection{Helicity 0 modes}

These channels are non-trivial solutions of the following equations:

\begin{equation}
X_{\sigma f}(x)=  \label{4.10A}
\end{equation}

\[
+g\frac{u_{f}^{2}(x)}{2E_{f}^{\prime }(x)-\Omega _{\sigma }(x)}A_{\sigma
}(x)+g\frac{g_{f}^{2}(x)}{2E_{f}^{\prime }(x)-\Omega _{\sigma }(x)}B_{\sigma
}(x) 
\]

\[
-g\frac{v_{f}^{2}(x)}{2E_{f}^{\prime }(x)-\Omega _{\sigma }(x)}C_{\sigma
}(x)-g\frac{u_{f}^{2}(x)}{2E_{f0}^{\prime }(x)-\Omega _{\sigma }(x)}%
D_{\sigma }(x) 
\]

\begin{equation}
Y_{\sigma f}(x)=  \label{4.10B}
\end{equation}

\[
g\frac{u_{f}^{2}(x)}{2E_{f}^{\prime }(x)+\Omega _{\sigma }(x)}C_{\sigma
}(x)+g\frac{v_{f}^{2}(x)}{2E_{f}^{\prime }(x)+\Omega _{\sigma }(x)}D_{\sigma
}(x) 
\]

\[
-g\frac{v_{f}^{2}(x)}{2E_{\sigma }^{\prime }(x)+\Omega _{\sigma }(x)}%
A_{\sigma }(x)-g\frac{u_{f}^{2}(x)}{2E_{f}^{\prime }(x)+\Omega _{\sigma }(x)}%
B_{\sigma }(x) 
\]

where

\begin{equation}
E_{f}^{\prime }(x)\equiv 2[E_{f}(x)-gu_{f}^{2}(x)v_{f}^{2}(x)]  \label{4.10C}
\end{equation}

and the definitions

\begin{equation}
\sum_{n}\int_{0}^{1}dx^{\prime }x^{\prime 2}X_{\sigma n}(x,x^{\prime
})u_{n}^{2}(x^{\prime })\equiv A_{\sigma }(x)  \label{4.11A}
\end{equation}

\begin{equation}
\sum_{n}\int_{0}^{1}dx^{\prime }x^{\prime 2}X_{\sigma n}(x,x^{\prime
})v_{n}^{2}(x^{\prime })\equiv B_{\sigma }(x)  \label{4.11B}
\end{equation}

\begin{equation}
\sum_{n}\int_{0}^{1}dx^{\prime }x^{\prime 2}Y_{\sigma n}(x,x^{\prime
})u_{n}^{2}(x^{\prime })\equiv C_{\sigma }(x)  \label{4.11C}
\end{equation}

\begin{equation}
\sum_{n}\int_{0}^{1}dx^{\prime }x^{\prime 2}Y_{\sigma n}(x,x^{\prime
})v_{n}^{2}(x^{\prime })\equiv D_{\sigma }(x)  \label{4.11D}
\end{equation}

Putting

\begin{equation}
u_{n}^{2}(x)+v_{n}^{2}(x)\equiv 1  \label{4.12A}
\end{equation}

\begin{equation}
u_{n}^{2}(x)-v_{n}^{2}(x)\equiv \alpha _{n}(x)  \label{4.12B}
\end{equation}

\begin{equation}
K_{\sigma }\equiv \frac{g}{2}(A_{\sigma }+B_{\sigma }-C_{\sigma }-D_{\sigma
})  \label{4.12C}
\end{equation}

\begin{equation}
K_{\sigma }^{\prime }\equiv \frac{g}{2}(A_{\sigma }-B_{\sigma }+C_{\sigma
}-D_{\sigma })  \label{4.12D}
\end{equation}

we find that

\begin{equation}
X_{\sigma f}(x)=\frac{g}{2\bar{E}_{f}^{\prime }(x)-\bar{\Omega}_{\sigma }(x)%
}[K_{\sigma }(x)+\frac{g}{2}\alpha _{f}(x)K_{\sigma }^{\prime }(x)] 
\label{4.13A}
\end{equation}

\begin{equation}
Y_{\sigma f}(x)=\frac{g}{2\bar{E}_{f}^{\prime }(x)+\bar{\Omega}_{\sigma }(x)%
}[-K_{\sigma }(x)+\frac{G}{2}\alpha _{f}(x)K_{\sigma }^{\prime }(x)] 
\label{4.13B}
\end{equation}

  This leads to the determinantal equation

\begin{equation}
S(\omega ,x)=\left| 
\begin{array}{ll}
S_{11}(\omega ,x) & S_{12}(\omega ,x) \\ 
S_{21}(\omega ,x) & S_{22}(\omega ,x))
\end{array}
\right| =0  \label{4.14}
\end{equation}

where

\begin{equation}
S_{11}(\omega ,x)=1-g\sum_{n} \mathcal{P} \int_{0}^{1}dx^{\prime }x^{\prime 2}\frac{%
2E_{n}^{\prime }(x^{\prime })}{(2\bar{E}_{n}^{\prime }(x^{\prime
}))^{2}-\omega (x)^{2}}  \label{4.15A}
\end{equation}

\begin{equation}
S_{12}(\omega ,x)=-\frac{g^{2}}{2}\sum_{n} \mathcal{P} \int_{0}^{1}dx^{\prime }x^{\prime
2}\frac{\alpha _{n}(x^{\prime })}{(2\bar{E}_{n}^{\prime }(x^{\prime
}))^{2}-\omega (x)^{2}}  \label{4.15B}
\end{equation}

\begin{equation}
S_{21}(\omega ,x)=2\omega \sum_{n} \mathcal{P} \int_{0}^{1}dx^{\prime }x^{\prime 2}\frac{%
\alpha _{n}(x^{\prime })}{(2\bar{E}_{n}^{\prime }(x^{\prime }))^{2}-\omega
(x)^{2}}  \label{4.15C}
\end{equation}

\begin{equation}
S_{22}(\omega ,x)=-1+g\sum_{n} \mathcal{P} \int_{0}^{1}dx^{\prime }x^{\prime 2}\frac{%
2E_{n}^{\prime }(x^{\prime })\alpha _{n}^{2}(x^{\prime })}{(2\bar{E}%
_{n}^{\prime }(x^{\prime }))^{2}-\omega (x)^{2}}  \label{4.15D}
\end{equation}

where $\mathcal{P}$ denotes the principal-value of the integral.

Again, an essential self-consistency condition for the solutions of these
equations is that the gap equations must be satisfied. Then it is trivial to
show that $\omega =0$ (``the Goldstone'') is a solution, if all parameters
that explicitly break the chiral-flavour symmetry of the effective
Hamiltonian are set to zero. This is evidently a well-known quantum effect.

Graphical solutions of eq(\ref{4.14}) \textit{for x =0} can be found from the
plots from Fig.\ref{fig.1} through Fig.\ref{fig.8} under various conditions. The zeroes of the
function $S(\omega ,0)$ can be interpreted as \textit{rest masses} of the
lowest boson-like vaccum excitations. All but two are situated amongst its
poles.

We consider three cases:

(i) A pure Goldstone mode, where

\begin{equation}
x_{u}=x_{d}=x_{s}=0  \label{4.16}
\end{equation}

In this case, our effective Hamiltonian is chiral - flavour invariant, but
this symmetry is broken by the vacuum. A common mass gap in the
quark/antiquark spectrum thus arises out of quantum effects.

The results in Fig.\ref{fig.1} and Fig.\ref{fig.2} refer to a toy model, but one that
correctly reproduces the bare bones of all our calculations, both without
self energy corrections and with self-energy corrections.
These figures illustrate a simplified case in which all q\={q} input energies in
Eqs(\ref{4.5A}) to (\ref{4.6B}) are of the form

\begin{equation}
\epsilon (i)=\epsilon \qquad i=1,2...,\Omega  \label{4.17}
\end{equation}

One can then calculate everything analytically. Fig.\ref{fig.1} illustrates how these
mass gaps and g are interrelated.We can thus uniquely renormalize g by
trading it with the in principle ''observable'' gap. The function $S(\omega )$
(Eq. \ref{4.14}) can thus be expressed as

\begin{equation}
S(\omega )=\omega^{2}R(\omega )  \label{4.18}
\end{equation}

with $R(0)\neq 0$. Fig.\ref{fig.2} shows both these functions. Besides the root at
zero (``the Goldstone'') there is another root, roughly 0.7 (arbitrary
units). The corresponding state vectors can easily be found from the above
relevant equations. The heavy state is made up of a maximally coherent
superposition of \textit{all} q\={q} basis states. The interplay between the single-particle
energies and the gap is decisive in determining in which frequency range the ``heavy''
actually shows up.

The remaining $\Omega -1$ q\={q} states are orthogonal to this one, and
essentially made up of a single q\={q} configuration ( i.e. some state i).

In Fig.\ref{fig.3} the calculations are repeated, but with the input single-particle
energy spectrum assumed in this paper.

The quarks energy-momentum relationships within our model space are
illustrated in Fig.\ref{fig.3} (again, with and without self-energy corrections): no
quark species remains massless, but the mass degeneracy is of course not
lifted. A feature worth noticing is the numerical influence of the
self-energy corrections on the input single particle/antiparticle spectrum.
Again, these are computed iteratively. About 10 iterations of the gap
variable are sufficient to reach satisfactory convergence.

In Fig.\ref{fig.4}a and b one can see the relevant branch of vacuum excitations. The quanta
of these excitations are boson-like. Note the presence of not only
Goldstones (lowest rest mass), but also heavy bosons with the largest rest mass, strongly
dependent both on the q,\={q} spectrum and on mass
gaps. Sandwiched in between the poles of $S(\omega)$, but not shown in Figs \ref{fig.4}a and b,
are other massive excitations with an intermediate degree of coherence necessary to make them
normalizable and mutually orthogonal. This basic structure of our results can be gleaned from
Fig.\ref{fig.2}.

(ii) A mixed Goldstone-Wigner mode, where

\begin{equation}
x_{u}=x_{d}=0\qquad ;\quad x_{s}>0  \label{4.19}
\end{equation}

In this case the quark degeneracy is obviously partially lifted. In Fig.\ref{fig.5}
and Fig.\ref{fig.6} one can see the results for a representative g that solves the gap equations.
The previous massless Goldstone now becomes a ``light'' pseudo-Goldstone. The move towards
``constituent masses'' for low momenta can still be seen.

  (iii) A mixed hierarquic Goldstone -Wigner mode, where

\begin{equation}
x_{d}>x_{u}>0\qquad ;\quad x_{s}\gg x_{d}  \label{4.20}
\end{equation}

The relevant results are shown in Figs. 7a,7b and 8.

Fig 7a shows an overall view of the energies-vs-momenta of the three species
of quarks (antiquarks). Fig7b gives a blow-up of this figure for very small
momenta. Note that the \textit{input} mass difference $x_{d}-x_{s}$ $\sim
0.0042$ has become \textit{reduced} in the \textit{output} rest energy
difference $E(d,0)-E(u,0)\sim 0.00078$. This result however does not take
into account virtual electroweak effects.

Finally, Fig.8 shows the corresponding results for bosonic excitation modes.
Again, the original Goldstone has been metamorphosed into a light
pseudo-Goldstone, which we now show together with all the heavies.
One recalls that our $%
\lambda =0$ basis is in fact a linear combination of spin $S=0$ and 
$S=1,\lambda =0$  and $I=0$ and $I=1, I_{3}=0$ states.
To identify these we need sources which inject these spins and isospins into
the vacuum. The trends are however unmistakable, even with such
relatively unsophisticated effective interactions as those adopted here.

We conjecture that this pseudo-Goldstone would be strongly coupled to the $%
\pi ^{0}$ whereas $\eta ^{0^{\prime }}$ would be strongly coupled to the
heavy coherent excitation that was previously mentioned (see App.B). The $%
\eta ^{0}$ would correspond to some intermediate superposition of the ''original'' q\={q}
background , but where it eventually would land strongly depends on
secondary residuals in the effective Hamiltonian not considered in this
paper.

\subsubsection{Helicity $\pm 1$ modes}

The equations determining excitations in these channels are formally the
same as (\ref{4.10A}- \ref{4.10C}). However, this requires a further parameter (g$_{1}$), and
thus we have not pursued the matter in this paper.

\section{Summary}

That there are self-consistent non-trivial stable solutions of the basic
equations (\ref{3.9}) and (\ref{4.14}) suggests that there is indeed support to the
conjecture \cite{10} that the quark vacuum may well exist in the same
universality class as the vacua (ground states) of some well-understood
superconducting Fermi liquid systems (e.g. $^{3}$He in its B-phase, \cite{10}).

If so, we have a well-understood effective field theory that could
reallistically be expected to provide support for the standard Chiral
Perturbation Theory of pions and etas \cite{2,3,4,5} in its attempts to account for
features that are not determined by chiral symmetry alone. In this
connection, we conjecture that Figs \ref{fig.4}, \ref{fig.6} and \ref{fig.8} show the rest mass spectrum
of boson-like particles, where the lowest mass will be strongly coupled to $%
\pi ^{0}$, the next lowest similarly strongly coupled to $\eta _{0}$ whereas
the heaviest (having a maximally coherent superposition of $u\bar{u},d\bar{d}
$ and $s\bar{s})$ could be the famous $\eta _{0}^{\prime }.$ This is a
robust feature of these calculations.

The gauge fields of the Standard Model play some sort of a background role
not analysed in this paper ; neither is the origin of the leading effective
interactions defining our model space (see however Ref. \cite{7}), clearly
controlling key features of the lightest mesons.

This conclusion confirms once again Dirac's belief of the fundamental
unity of Physics \cite{6}.

\ack
I would like to thank Geoff Oades for his patient advice, friendly criticism
and much practical help.

\appendix

\section{ }

We sketch here for the benefit of the reader not familiar with
the contents of a ``Field Theory without dead wood'', the streamlined
version of Field Theory proposed by Dirac \cite{6}, as the most convenient tool
for expressing our ideas.

Given the effective Hamiltonian, our task is to solve the Heisenberg
equations of motion \cite{6}:

\begin{equation}
i\frac{\partial }{\partial t}\hat{K}(t)=[\hat{H},\hat{K}(t)]  \label{AI.1}
\end{equation}

for q-numbers that we presume to be constants of motion $\hat{K}(t)$.

We start our approximation scheme by partitioning the Hamiltonian :

\begin{equation}
\hat{H}=\hat{H}_{0}^{\prime }+\hat{V}^{\prime }  \label{AI.2}
\end{equation}

We first diagonalize $\hat{H}_{0}^{\prime }$ as best as we can. It may
however break some of the symmetries of $\hat{H}$. Approximation schemes for
finding stationary solutions of (\ref{AI.1}) must therefore be designed so that
they respect the conservation laws associated with these symmetries to any
desired accuracy.

Introducing q-numbers in Dirac Picture

\begin{equation}
\hat{O}_{D}(t)=e^{i \, \hat{H}_{o}^{\prime } \, t} \, \hat{O}(t)%
\, e^{-i \, \hat{H}_{o}^{\prime } \, t}  \label{AI.3A}
\end{equation}

\begin{equation}
\hat{V}_{D}(t)= \, e^{i \, \hat{H}_{o}^{\prime } \, t} \,%
\hat{V}^{\prime } \, e^{-i \, \hat{H}_{o}^{\prime } \, t} 
\label{AI.3B}
\end{equation}

we then obtain:

\begin{equation}
i\frac{\partial }{\partial t}\hat{O}_{D}(t)=[\hat{V}_{D}(t),\hat{O}_{D}(t)] 
\label{AI.4}
\end{equation}

Setting

\begin{equation}
\hat{O}_{D}(t)=\sum_{n=0}^{\infty }\hat{O}_{D}^{\left( n\right) }(t) 
\label{AI.5}
\end{equation}

we find the conditions:

\begin{equation}
\frac{\partial }{\partial t}\hat{O}_{D}^{\left( 0\right) }(t) \, =0 
\label{AI.6}
\end{equation}

and the recursion relations

\begin{equation}
i\frac{\partial }{\partial t}\hat{O}_{D}^{\left( n\right) }\left( t\right) =[%
\hat{V}_{D}(t),\hat{O}_{D}^{\left( n-1\right) }(t)]\qquad n\geq 1  \label{AI.7}
\end{equation}

   The first condition gives

\begin{equation}
\hat{O}_{D}^{\left( 0\right) }(t)\equiv \hat{K}  \label{AI.8}
\end{equation}

where the q-number $\hat{K}$ is not explicitly time-dependent.

The remaining corrections are given by the recursion relation :

\begin{equation}
\hat{O}_{D}^{\left( n\right) }(t)=\left( -i\right) \int^{t}dt_{1}[\hat{V}%
_{D}(t_{1}),\hat{O}_{D}^{\left( n-1\right) }(t_{1})]\qquad n\geq 1 
\label{AI.9}
\end{equation}

Before one can interpret the results at any given stage, one must apply the
q-numbers thus obtained to the standard (Heisenberg) \textit{reference ket} $%
|0>_{t}$. This reference ket is \textit{not} the usual ``vacuum state
vector'', as we have no such concept here, but simply helps to define normal
products \textit{at time t only}. We adopt the following procedure \cite{6}:

(i) Compute the commutator and then apply Wick's theorem \cite{8}, using of
course all the previously established relevant conditions, if any.This makes
significant the need for normal ordering in this q-number theory;

(ii) Carry out the time integrations;

(iii) Pick up the coefficient of the i'th term , and interpret this
coefficient as the \textit{prediction/postdiction }of this theory for what
the \textit{amplitude }for the i'th configuration (state) to emerge \textit{%
at any time t} is, i.e.:

\begin{equation}
\hat{O}(t)|0>_{t}=\sum_{N=0}^{\infty }\hat{O}^{(N)}(t)|0>_{t}=\sum_{N=0}^{%
\infty }\sum_{\nu }\{|\nu ><\nu |\hat{O}^{(N)}(t)|0>\}_{t}  \label{AI.10}
\end{equation}

As emphasized by Dirac, it is not possible within this framework to make
predictions, or postdictions, of what can be, or could be, observed at any
time other than t, even if one possesed a complete knowledge of all
amplitudes at time t. A critical discussion and assesment of this procedure
was given by Dirac in the quoted references.

\section{ }

Let us speculate on the structure of the lightest self-conjugate physical
mesons at rest. We rely on an assertion, common to all versions of the Quark
Model, that mesons are q\={q} colour singlet composites . So, we define
these as the composites

\begin{eqnarray}
\mathbf{\hat{O}}\equiv \mathbf{\Gamma }_{\nu IJ\lambda }^{+}(\vec{p}%
)&=&\sum_{i}\int_{\mathcal{D}}d^{3}\vec{p}^{\prime }X_{\nu (iI)J\lambda
}(p,p^{\prime })\Gamma _{(iI)J\lambda }^{+}(\vec{p}^{\prime }) \nonumber \\
& & -\sum_{i}\int_{%
\mathcal{D}}d^{3}\vec{p}^{\prime }Y_{\nu (iI)J\lambda }(p,p^{\prime })\Gamma
_{(iI)J\lambda }(\vec{p}^{\prime })  \label{AII.1}
\end{eqnarray}

  Putting

\[
\nu IJ\lambda \equiv M 
\]

\[
(iq)\lambda _{1}\lambda _{2}\equiv r\ ;\qquad \lambda =\lambda _{1}-\lambda
_{2} 
\]

we get

\begin{equation}
\mathbf{\hat{O}}\equiv \mathbf{\Gamma }_{M}^{+}(\vec{p})=\sum_{r}\int_{%
\mathcal{D}}d^{3}\vec{p}^{\prime }X_{Mr}(p,p^{\prime })\Gamma _{r}^{+}(\vec{p%
})-\sum_{r}\int_{\mathcal{D}}d^{3}\vec{p}^{\prime }Y_{Mr}(p,p^{\prime
})\Gamma _{r}(\vec{p})  \label{AII.2}
\end{equation}

where

\begin{equation}
X_{\nu (iI)J\lambda }(p)\equiv \sum_{q=-i}^{+i}\sum_{\lambda _{1}\lambda
_{2}}(\frac{1}{2}\lambda _{1}\frac{1}{2}-\lambda _{2})|J\lambda
)(iqi-q|I0)X_{M(iq)\lambda _{1}\lambda _{2}}(p)  \label{AII.3}
\end{equation}

and \textit{mutatis mutandis }for the Y amplitudes.

The amplitudes $X_{\nu (iI)J\lambda }(p,p^{\prime })$ and $Y_{\nu
(iI)J\lambda }(p,p^{\prime })$can be determined after finding non-trivial
solutions of the system of equations (\ref{3.21A}) and (\ref{3.21B}).

Meson-like particles carrying spin J and isospin I at rest would then be
describable by wave-packets such as

\begin{equation}
\Gamma _{M}^{+}=\int d^{3}\vec{p}\Psi _{M}(\vec{p})\Gamma _{M}^{+}(\vec{p}) 
\label{AII.4}
\end{equation}

where

\begin{equation}
\int d^{3}\vec{p}|\Psi _{M}(\vec{p})|^{2}=1  \label{AII.5}
\end{equation}

Within these approximations , these composites would appear to be
structureless point-like spin 0 and spin 1 \textit{bona fide }free bosons
carrying good quantum numbers, in all observations involving small momentum
transfers.The amplitudes X and Y (\ref{AII.3}) will be explicitly needed, should
one wish to further test this conjecture, e.g. by studying electroweak
probings of these point-like mesonic quasiparticles at rest.Some residual 
\textit{strong interaction} effects, describable within the present scheme,
should be treated by first returning to our basic theory (Section 2) and
going beyond the leading approximations, but still within the framework of
conserving approximations. In any case, this procedure could be a good
support to Chiral Perturbation theory as applied to pions and etas \cite{3}
beyond its leading approximations.

Specific, but easily handled, field theoretical models such as this one
could, among other duties, thus help in testing the widespread suspicion
that the non-chiral quark model description of the lightest mesons, as used
in more or less sophisticated so-called relativised versions,e.g. \cite{9}, is at
best too crude to account for these particular mesons and their interactions.

\Figures

\begin{figure}

\caption{\label{fig.1}
This shows how the ``coupling constant g'' is renormalized in
this paper,i.e. how it can be uniquely traded with an (in principle)
``observable'' mass-gap parameter $\Delta $. g is dimensionless, and the gap
parameter is in units of 1 GeV. The lower curve a) does \textit{not }include
self-energy corrections, but the upper curve b) does.

The parameters used in this toy model are: $\epsilon =0.2$ ;$\Omega =$ $%
1000.$}

\end{figure}

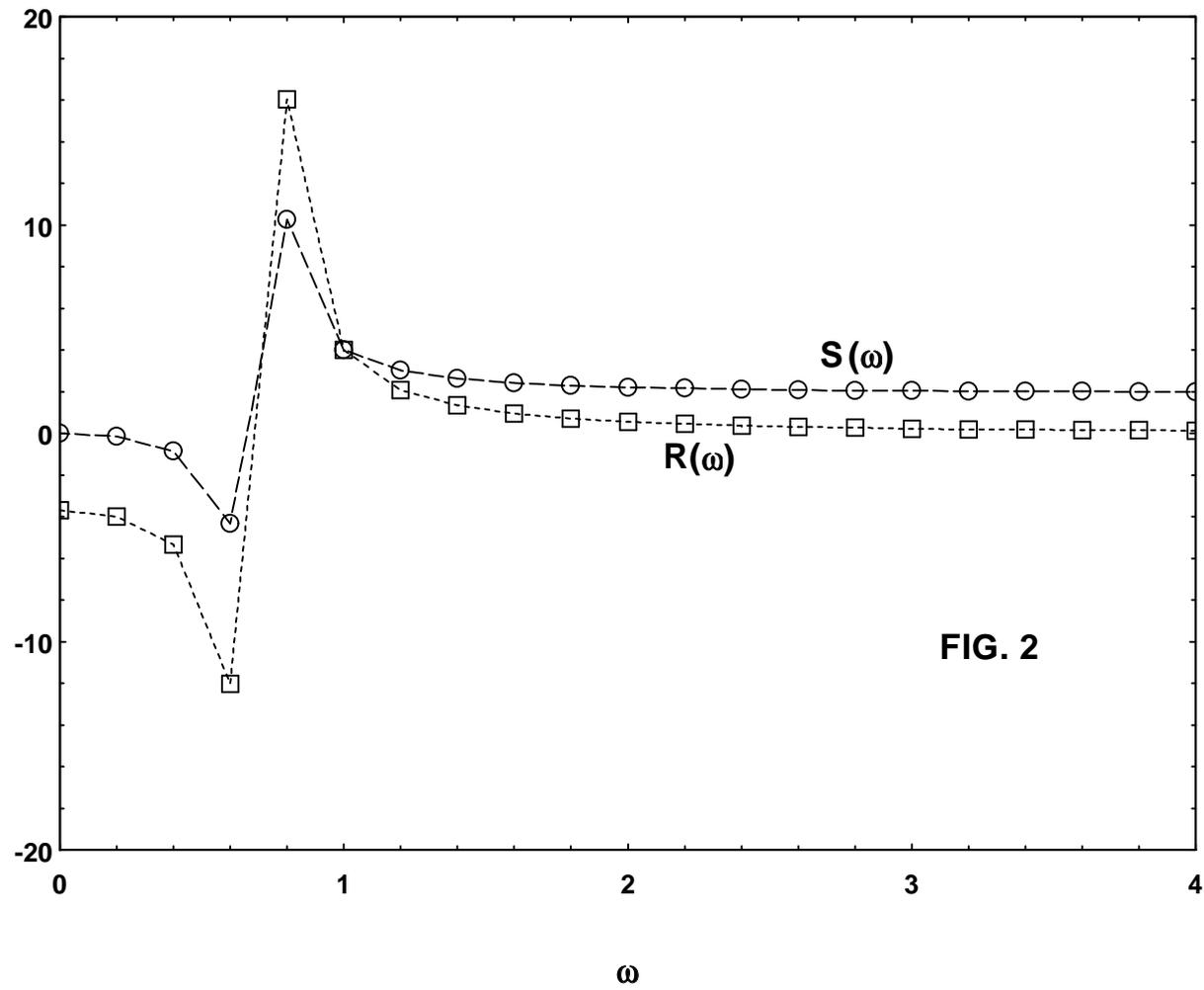
\begin{figure}

\caption{\label{fig.2}This illustrates how certain relevant vacuum Bose modes arise
in the toy model : a Goldstone (playing the role of pions) and a heavy mode
(playing the role of $\eta ^{0\prime }$ ?). This is a highly coherent mode,
in contrast to the other modes, wich remain in nearly the same state as for
$g=0$, and are sandwiched among the poles of $S(\omega).$ The parameters
used in making Fig2. are: $\epsilon =0.2,\Delta =0.3$.}

\end{figure}

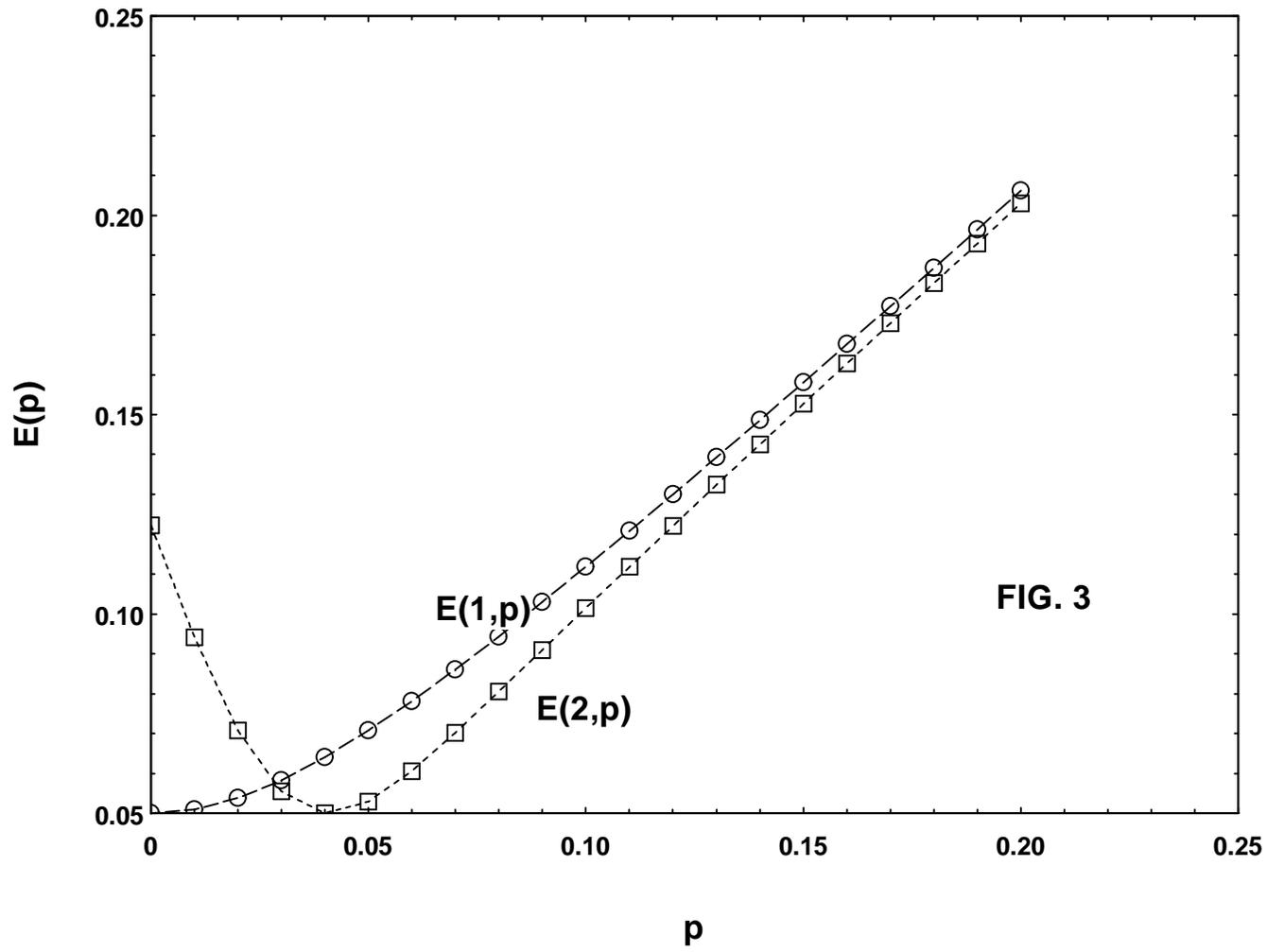
\begin{figure}

\caption{\label{fig.3}We plot here the energy - momentum relationship in case A
(mass-degenerate quarks) . Dimensionsful variables are in units of 1 GeV. No
self-energy corrections are included in curve $E(1,p)$ , but they are
fully included in the curve $E(2,p)$. Note the importance (in our case) of
the self-energy corrections that tend to push the $p=0$ masses towards
something like ''observed constituent '' values, especially for small
momenta.
The parameters used in making Fig.3 and Fig.4 are: $\Delta =0.05\
(g=1.34).$}

\end{figure}

\begin{figure}

\caption{\label{fig.4}a) We plot here the function $S(\omega ,0)$ in case A (massless
quarks) . Variables having dimensions are in units of 1GeV. We show only the zeros of
$S(\omega)$  which are of interest from the perspective of this paper, i.e. the
lightest and the heaviest of the excitations. Note the Goldstone at $\omega =0$.

b) Note the ''heaviest'', highly coherent excitation at
about $\omega \sim 2GeV.$ The self-consistency conditions are fully
respected. Intermediate zeroes have not been shown.}

\end{figure}

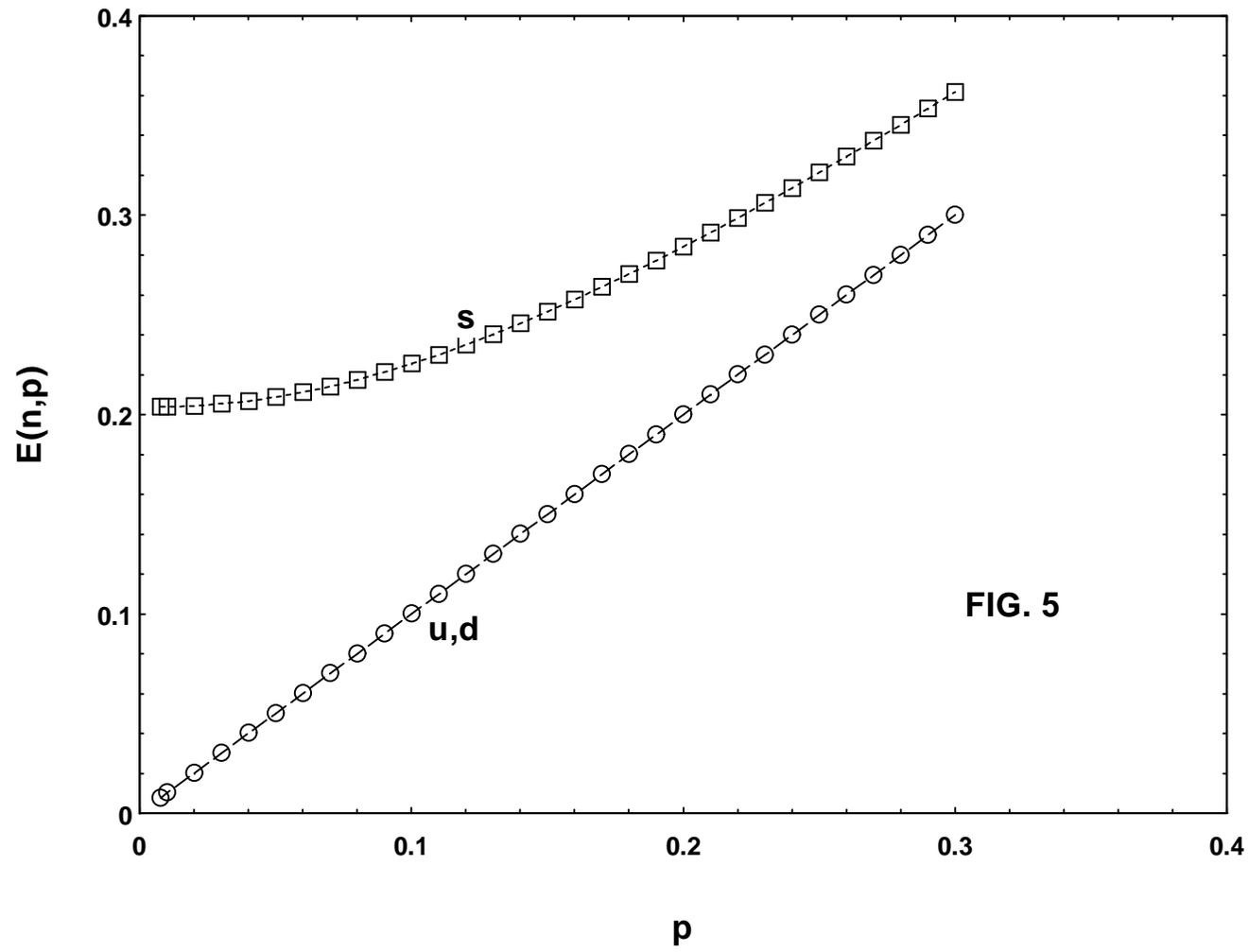
\begin{figure}

\caption{\label{fig.5}We plot here the energy - momentum relationship in case B
(massless u,d- quarks) . Variables with dimensions are in units of 1 GeV.
Self-energy corrections are fully included.

The parameters used in making this figure are: $x_{u}=x_{d}=0$; $x_{s}=0.2$
and $g=0.08.$

We find that $E(u,d;0)=0.008$ ; $E(s;0)=0.205.$}

\end{figure}

\begin{figure}

\caption{\label{fig.6}We plot here the function $S(\omega ,0)$ in case B.
Quantities with dimensions are in units of 1 GeV. Self-energy
corrections are fully included. The same parameters as in Fig5.

Note the disappearance of the Goldstone ($S(0,0)\sim -0.9$).Instead, we get
a pseudo-Goldstone with a rest mass somewhere between $\omega =0.4$ and $%
\omega =0.6.$

The ``heavy'' of special interest to us is the last zero. Its rest mass
emerges between $\omega =1.2$ and$\omega =1.4$. with our particular choice
of parameters. In all cases the details of the q\={q} spectrum and the gap
parameters are crucial in determining where the masses actually emerge.}

\end{figure}

\begin{figure}

\caption{\label{fig.7}a) We plot here the energy - momentum relationship in case C ( 3
massive quarks) . The dimmensionfull variables are in units of 1 GeV.
Self-energy corrections are fully included. The parameters used in this
calculation are:

$x_{u}=$ $0.0051,\ x_{d}=0.0093,\ x_{s}=0.1750\qquad $; $g=0.05$

b) We show in this figure for clarity a blow-up of the low
momenta part of a)}

\end{figure}

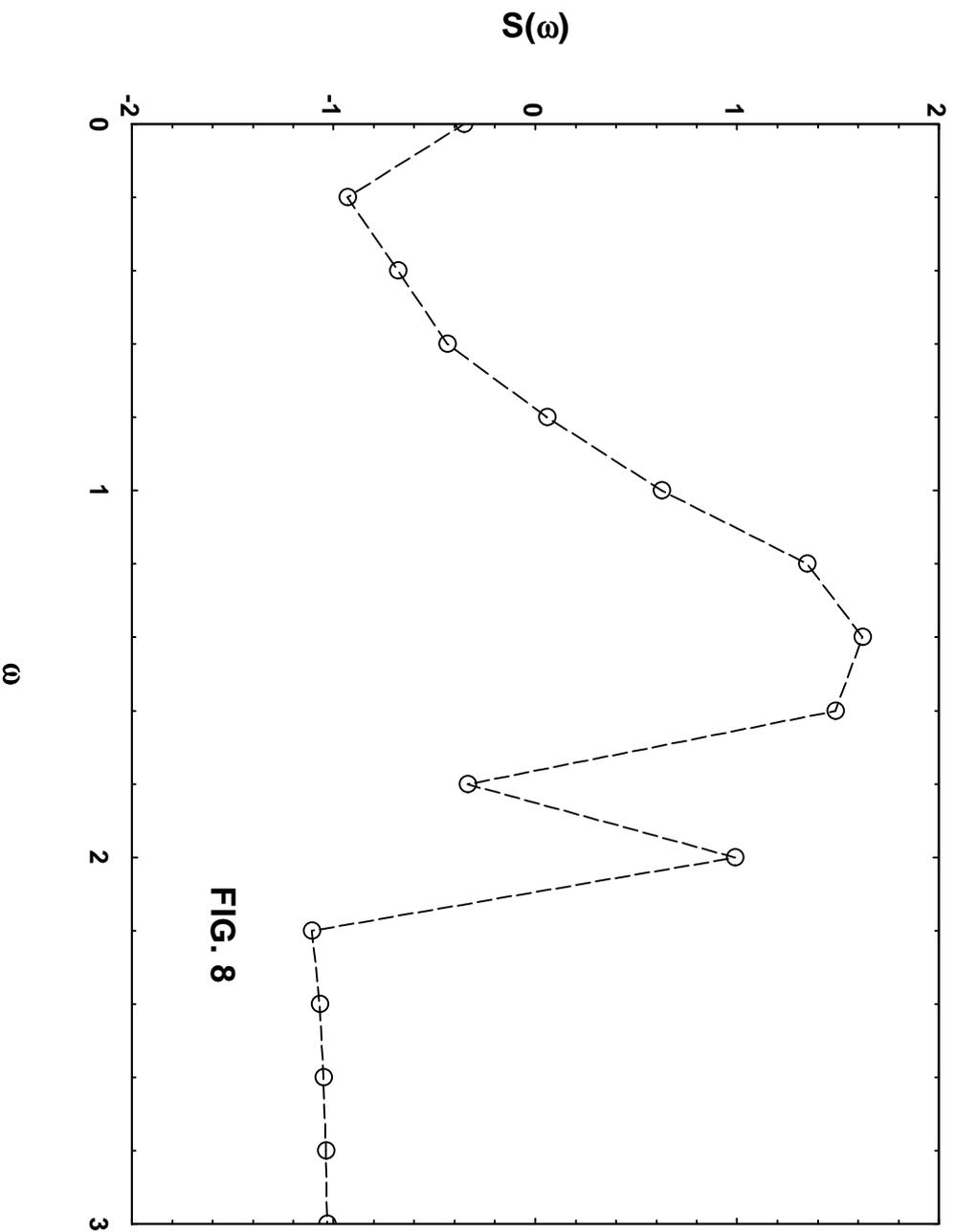
\begin{figure}

\caption{\label{fig.8}We plot here the function $S(\omega ,0)$ in case C (massive
quarks) . Dimensionsful variables are in units of 1 GeV. Only energy
corrections on the q,\={q} energies consistent with the often mentioned
self-consistency conditions are fully included. Same parameters as used in
Fig.7a. See text for further comments.}

\end{figure}

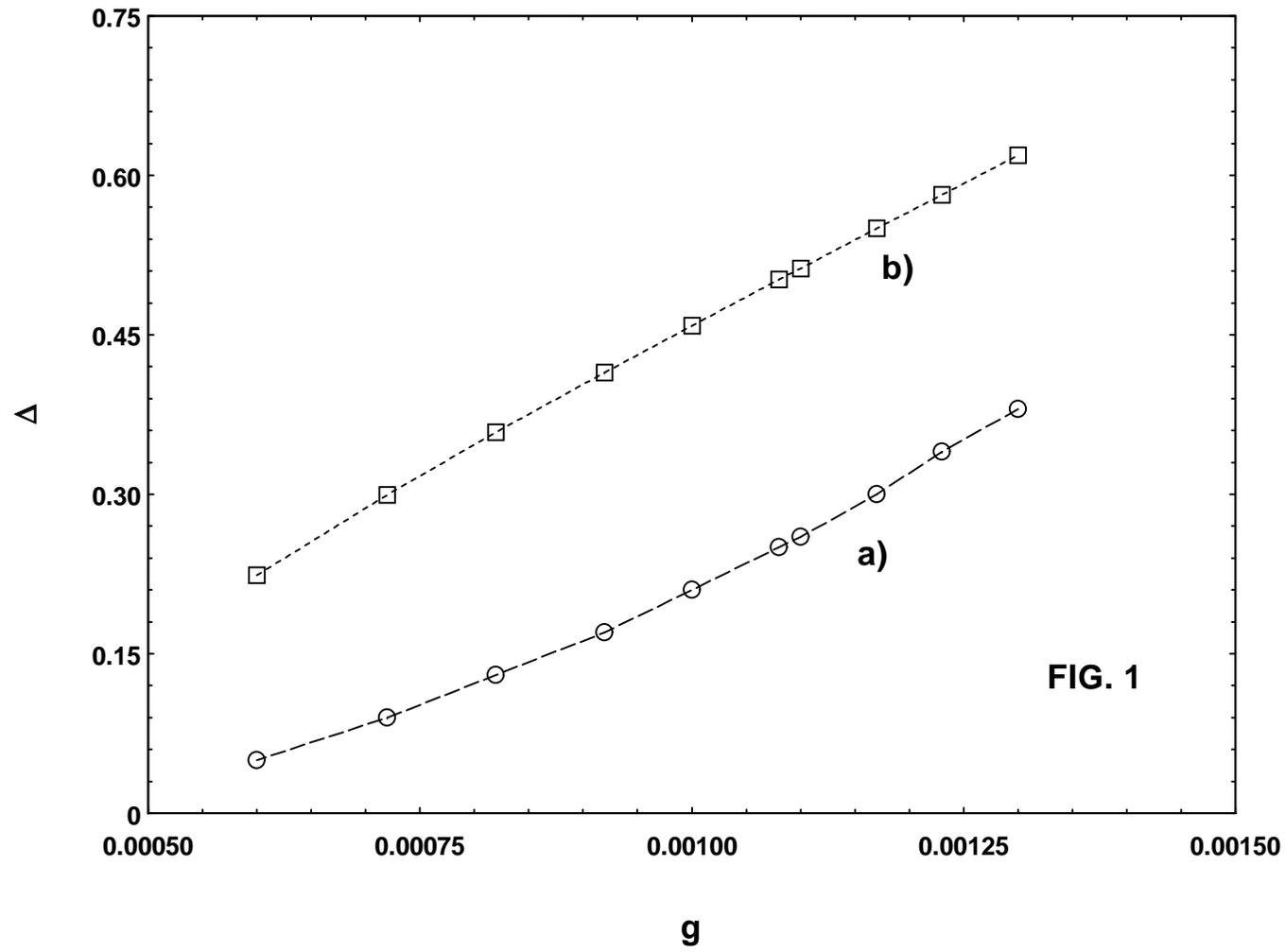
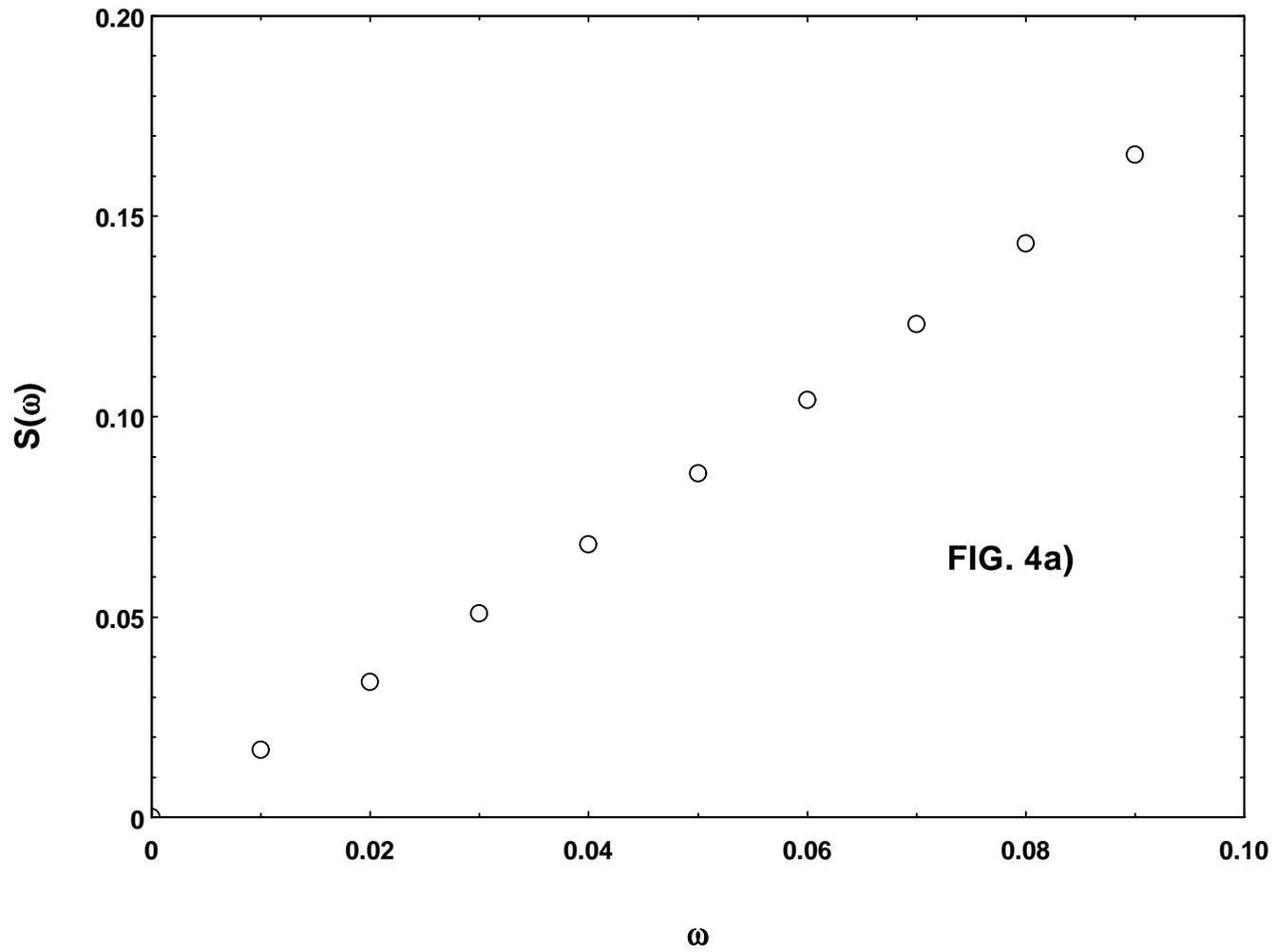
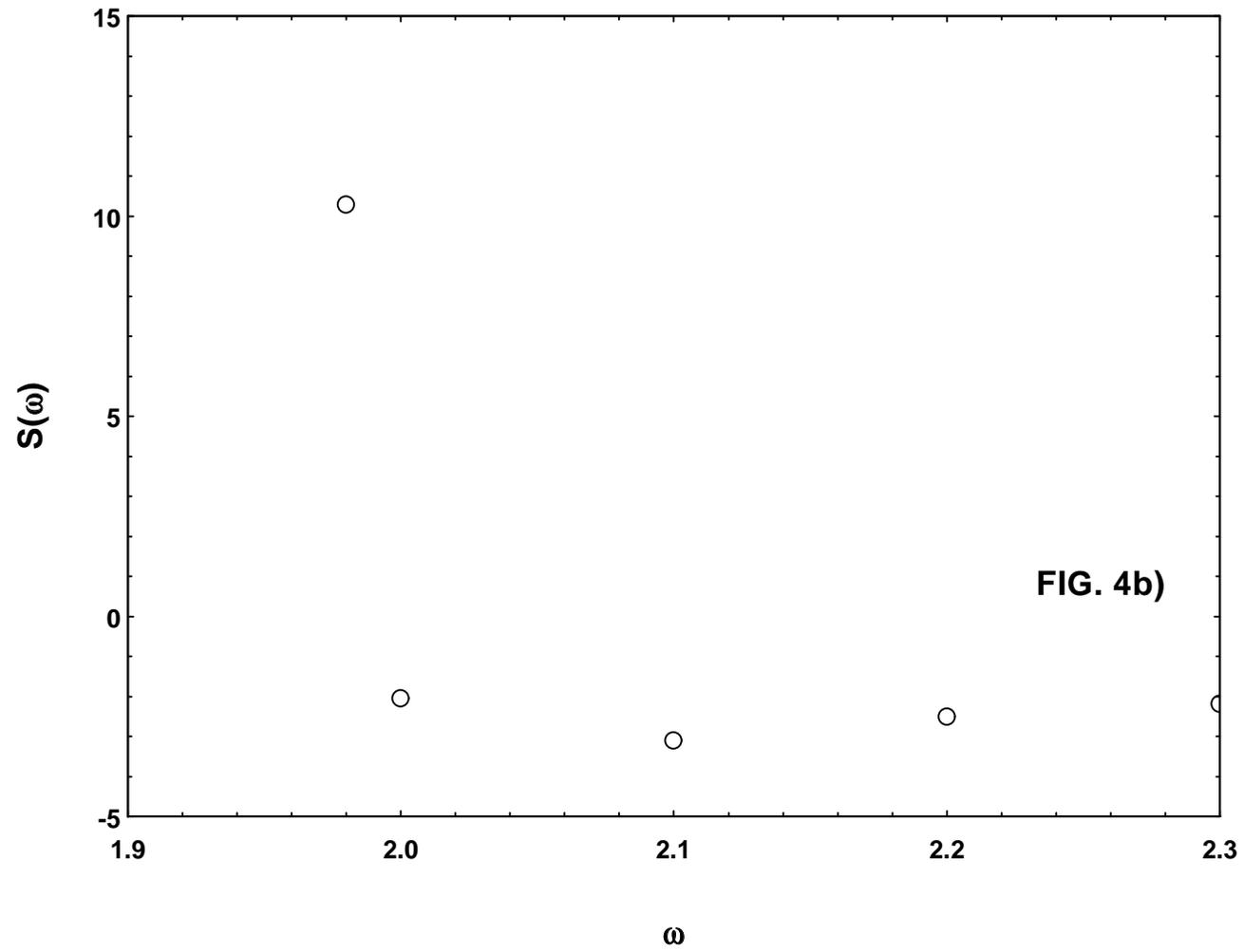
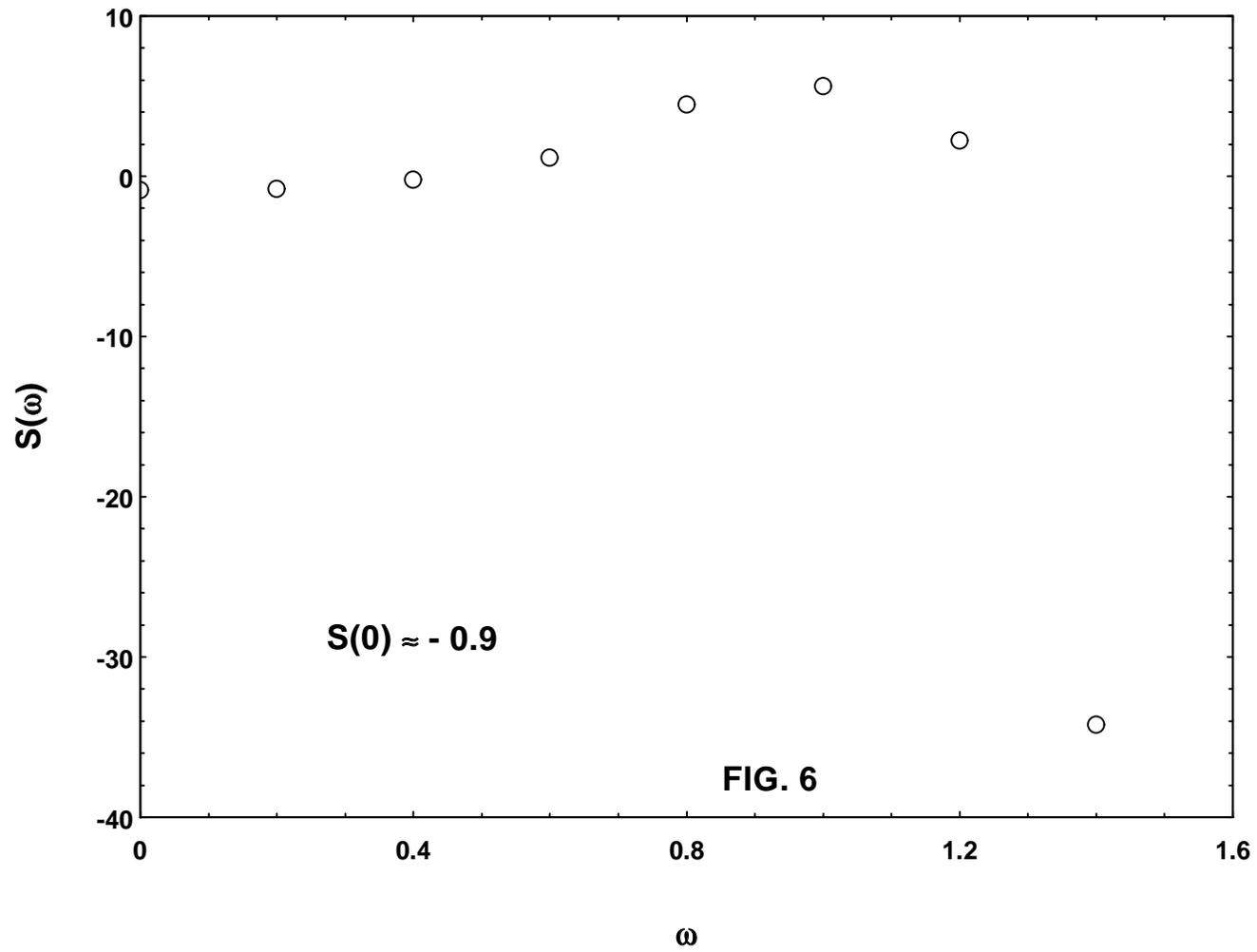
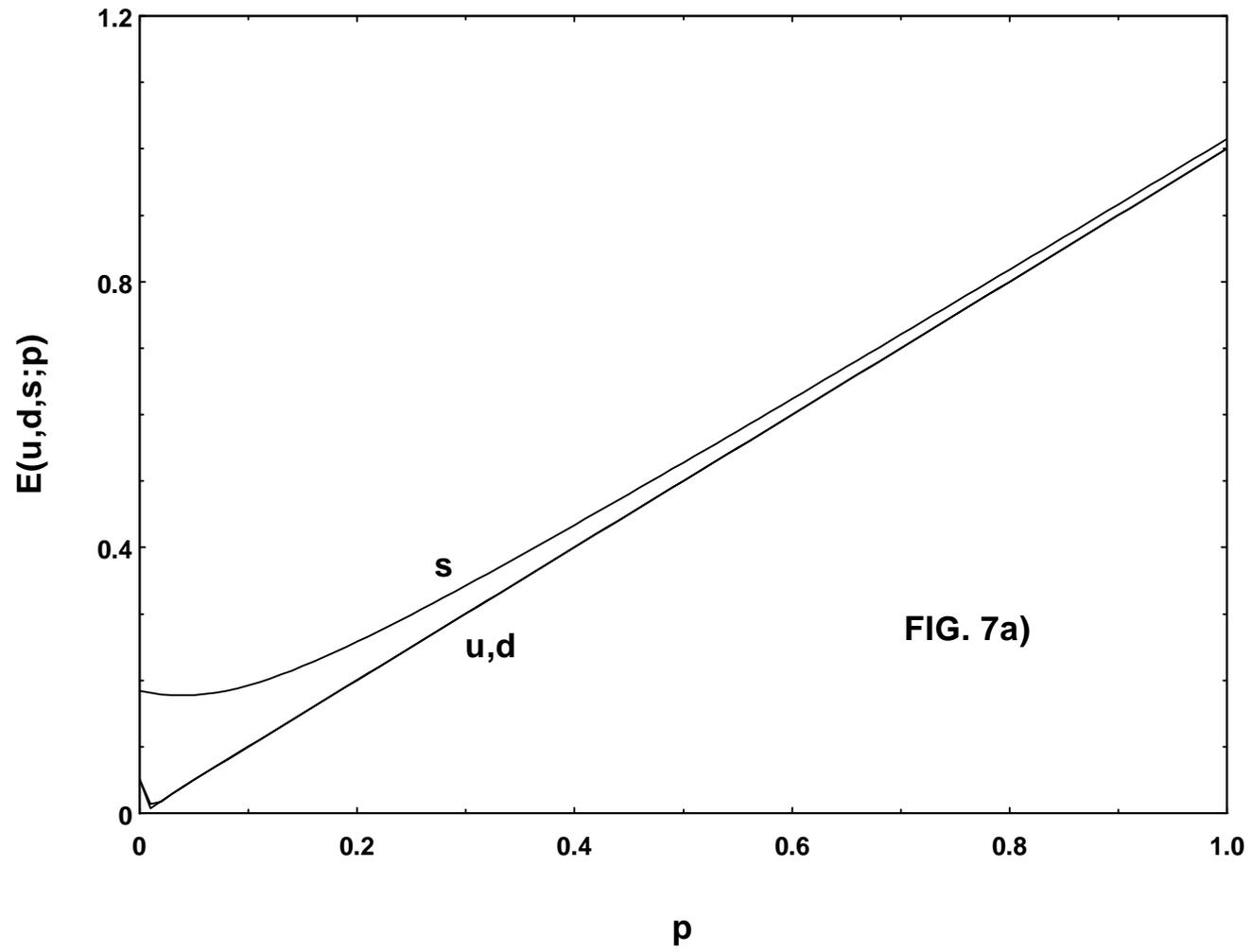
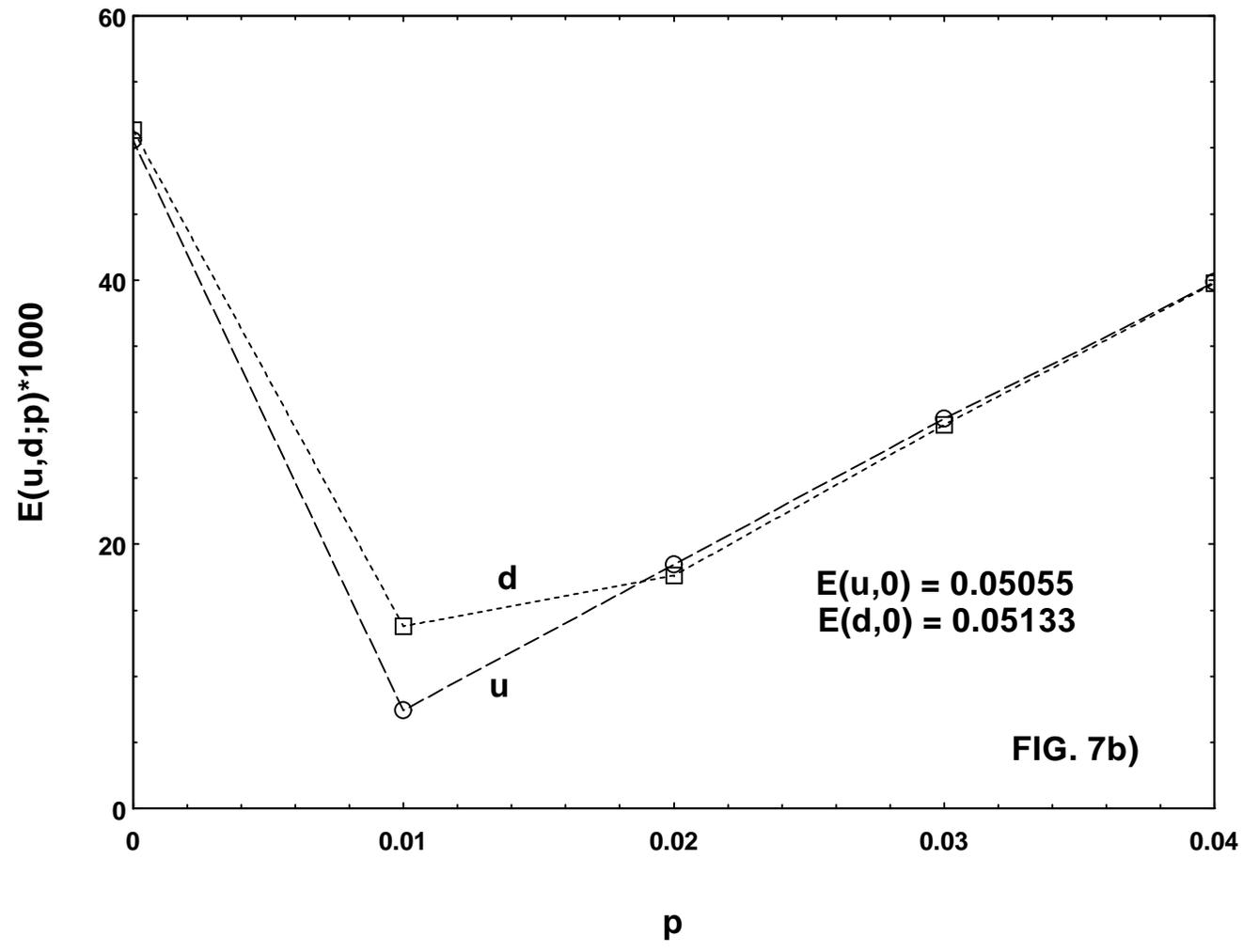
\end{document}